\documentclass[twocolumn,showpacs]{revtex4}
\usepackage{graphicx}
\usepackage{epsfig}
\usepackage{amsmath}
\usepackage{amssymb}

\newcommand{\beq}{\begin{eqnarray}}
\newcommand{\eeq}{\end{eqnarray}}

\begin{document}

\title{Ferroelectricity due to orbital ordering in E-type undoped rare-earth manganites.}

\author{Paolo Barone}
\affiliation{CNR-SPIN, 67100 L'Aquila, Italy}
\author{Kunihiko Yamauchi}
\affiliation{CNR-SPIN, 67100 L'Aquila, Italy}
\author{Silvia Picozzi}
\affiliation{CNR-SPIN, 67100 L'Aquila, Italy}

\begin{abstract}
Aiming at understanding the origin of the electronic contribution to ferroelectric polarization in undoped manganites,
we evaluate the Berry phase of orbital-polarizable Bloch electrons as an orbital ordering (OO) establishes in the
background of an antiferromagnetic E-type configuration. The onset of OO is tuned by the Jahn-Teller (JT) interaction
in a tight-binding model for interacting electrons moving along zigzag chains. A finite polarization is found as soon as
the JT coupling is strong enough to induce OO, supporting the large electronic contribution predicted from first principles.
\end{abstract}

\maketitle

The coexistence of ferroelectrically polarized state with long-range magnetic order is usually referred to as multiferroic
behavior\cite{picozzi.rev}. Among other multiferroic materials,
orthorhombic rare-earth manganites $R$MnO$_3$ ($R$ belonging to lanthanum series) represent an important class
of ``improper multiferroics''\cite{cheong}, where the ferroelectric polarized state
is not only coexisting with, but also intrinsically related to some kind of magnetic order. Several degrees of freedom
(spin, charge, orbital, lattice) are active in these systems and are responsible for their very rich phenomenology. Starting
from $R=$La and moving along the series to smaller ionic radius $r_R$, the ground state changes from
antiferromagnetic A-type (AFM-A) to E-type (AFM-E), through an incommensurate spiral
structure for $R=$Tb, Dy\cite{kimura}. The incommensurate magnetic phase of TbMnO$_3$ and DyMnO$_3$ is responsible for the
observed weak ferroelectric polarization\cite{kimura2}, which is ultimately driven by spin-orbit coupling via the 
Dzyaloshinskii-Moriya interaction\cite{mostovoy,dagotto1}. The relativistic origin of ferroelectricity in these compounds
is reflected in the weak $P$ (smaller than $0.1\mu C/cm^2$\cite{kimura2}). On the other hand, 
also collinear AFM-E magnetic order has been predicted to induce a ferroelectric polarization via an
exchange-striction mechanism\cite{dagotto2}, where the double-exchange interaction between Mn $e_g^1$ electrons in the
symmetry-broken $\uparrow \uparrow \downarrow \downarrow$ spin configuration is responsible for polar atomic displacements
of oxygens bridging Mn ions. In a following paper\cite{picozzi}, this picture has been partially confirmed by means of
density-functional theory (DFT) calculations, reporting $P\sim 6\mu C/cm^2$ for $o-$HoMnO$_3$.
The DFT analysis also pointed out
that a comparable contribution to the total polarization has a purely quantum mechanical origin,
possibly related to the AFM-E-induced asymmetric electron hopping of orbitally polarized $e_g^1$ states. The microscopic origin of such
electronic contribution to $P$ has been provided in terms of maximally localized Wannier
functions (WFs)\cite{yamauchi}. The DFT study showed that WF centers in all 
AFM-E $o-R$MnO$_3$ ($R=$Ho, Er, Tm, Lu),
are largely displaced with respect to corresponding ionic positions; thus, the total
$P$ arises from the sum of an ionic contribution (driven by the exchange-striction
mechanism) and a purely electronic one\cite{yamauchi}.

From the experimental point of view, only a small polarization ($P\sim0.01\mu C/cm^2$) has been reported for
$o-$HoMnO$_3$\cite{lorenz}, at odds with theoretical prediction.
Only recently experimental results for AFM-E $o-$TmMnO$_3$ suggested that $P$ could actually pass the $1\mu C/cm^2$
threshold\cite{yu}. By the way, we notice that the thermodynamically stable phase for $R$MnO$_3$  with
$r_R$ smaller than that of Dy is hexagonal rather than orthorhombic\cite{goto}: the desired
perovskite structure can be obtained, e.g., by high-pressure synthesis, leading however to polycrystalline samples that
can limit the accuracy in  measuring the predicted properties.

\begin{figure}[t]
\includegraphics[width=8cm]{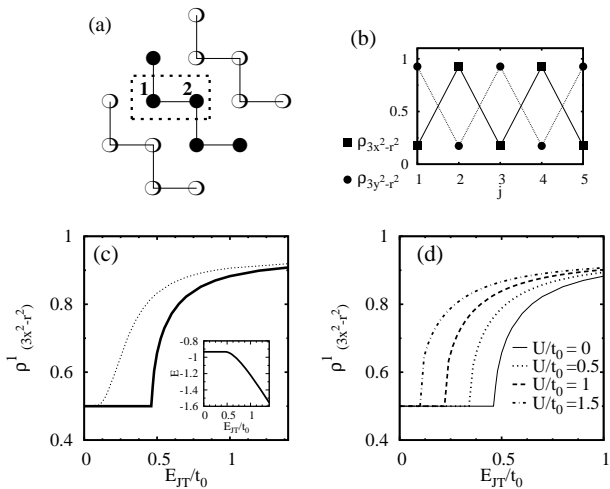}
\caption{(a) Unit cell for the zigzag ferromagnetic chain in the AFM-E phase: the hopping direction changes
periodically within the chain as $\{..x,y,x,y..\}$. (b) Orbital densities
$\rho_{3x^2-r^2}, \rho_{3y^2-r^2}$ as a function of position within a chain at $E_{JT}=1$, when OO is almost fully
developed. (c) Average orbital density for $d_{3x^2-r^2}$-state at site 1 as a function of JT interaction (inset: total energy
smooth evolution). Dotted line is the same
quantity evaluated in a linear chain, when $\varphi_{\vert i-j\vert}=const$. (d) Same as (c), but for different
values of interorbital Coulomb interaction.}\label{fig1}
\end{figure}

Aim of this Letter is to set a clear correspondence between the electronic polarization and the onset of orbital ordering
(OO) on the
background of the magnetic AFM-E configuration. The proposed picture is completely new in the field of improper
multiferroicity and relies on the orbital polarizability of electrons on a specific
magnetic background. We will introduce a simple
model which is expected to reproduce
the general features of the ground state of AFM-E orbital-ordered $R$MnO$_3$. The AFM-E configuration of
$t_{2g}$ spins is treated as a
constraint on the $e_g$ electrons motion, a reasonable assumption in the limit of very large Hund coupling.
Its stability has been already discussed, e.g. in Refs.
[\onlinecite{kimura}],[\onlinecite{yamauchi}], where the competition between kinetic energy (double exchange) and
superexchange interactions
between the Mn $t_{2g}$ spins is discussed as a function of the ionic radius $r_R$ (as pointed out recently, the Jahn-Teller
(JT)
interaction also may play a relevant role\cite{kumar}).
In limit of infinite Hund coupling,  $e_g$ electrons can only
hop between sites with ferromagnetically aligned core-spins; this implies that hopping processes in the AFM-E phase
are allowed only within
one-dimensional zigzag chains of parallel core spins [Fig.\ref{fig1}(a)]. OO may be stabilized by an electron-lattice
JT interaction, which causes also the Bloch electrons within the chains to acquire a geometric
Berry phase arising from a conical intersection of the potential energy surfaces\cite{koizumi}; we stress the fact,
however, that also a correlation-mediated mechanism may stabilize OO\cite{kugel}.
In the framework of the modern theory of polarization, the appearance of this Berry phase can
give rise to a polarization whose origin is purely quantum mechanical, being intimately connected to a current flow
inside the bulk\cite{king-smith}. We will show that this is indeed the case, but that the geometric phase of the
OO state is actually not enough, being the interplay with the specific
topology induced by the underlying magnetic background the boost for ferroelectric polarization.

Let us introduce the Hamiltonian which describes the motion of interacting $e_g$ electrons within zigzag chains:
\beq\label{ham}
H=&-&\sum_{\langle jl\rangle\gamma\gamma'}\, t_{\gamma\gamma'}\,a^{\dagger}_{\gamma j}a^{\phantom{\dagger}}_{\gamma' l}\,+\,
U\,\sum_j \, n_{\alpha j}\,n_{\beta j} 
 \\ \, &+& E_{JT}\,\sum_j\,
\left[\,2\,(\,q_{2j}\tau_{xj}\,+\,q_{3j}\tau_{zj}\,)+q_{2j}^2+q_{3j}^2 \right]. \nonumber
\eeq
The first term describes the electron transfer between nearest-neighbor Mn sites,
where $a_{\gamma j}^\dagger$ 
creates a particle at site $j$ in electronic states stemming from Mn$^{+3}$ orbitals $d_{x^2-y^2} (\alpha)$ and
$d_{3z^2-r^2} (\beta)$.
The hopping amplitudes are $t_{\alpha\alpha}=3t_0/4,\,t_{\beta\beta}=t_0/4$ and $t_{\alpha\beta}=\pm\sqrt{3}t_0/4$
\cite{slater}, where
$t_0=(pd\sigma)^2$ is the
energy unit and the sign appearing in $t_{\alpha\beta}$ depends on the hopping direction
along the zigzag chain (i.e. $t_{\alpha\beta}^x=-t_{\alpha\beta}^y$). As we will see, this implies that at each site
electrons gain a phase
that depends on the orbital through which they pass. 
The second term describes the interorbital interaction, the only Coulomb interaction that is left between
$e_g$ electrons in the infinite Hund coupling limit, while the last term accounts for the JT interaction
with dimensionless $q_{\nu j}\,=(k/g)\,Q_{\nu j}$, where $Q_{\nu j}$ are the
Jahn-Teller-active modes\cite{kanamori}, $k,g$ are respectively the elastic JT stiffness and the bare JT coupling,
and
$E_{JT}=g^2/(2k)$ is the static JT energy.  $\tau_{\mu
j}=\sum_{\gamma\gamma'}a_{\gamma j}^\dagger\,\sigma_{\gamma\gamma'}^{\mu}\,a_{\gamma' j
}^{\phantom{\dagger}}$ is the orbital pseudospin, being $\sigma_{\gamma\gamma'}^{\mu}$ the Pauli matrices,
while $n_{\gamma
j}=a_{\gamma j}^{\dagger}a_{\gamma j}^{\phantom{\dagger}}$ are orbital density operators. We treat the on-site
correlation in a mean-field framework,
by linearizing it in the optimum local basis in order to keep track of the orbital degrees of
freedom\cite{dagotto.rev}, finding $n_{\alpha j}n_{\beta j}\approx (2 n_j\langle n_j\rangle -2\tau_{z j}\langle \tau_{z j}\rangle
-2\tau_{x j} \langle \tau_{x j}\rangle -\langle n_j\rangle^2 + \langle \tau_{z j}\rangle^2 + \langle \tau_{x
j}\rangle^2)/4$.

As the phase change in hopping amplitudes plays a relevant role in establishing a ferroelectric polarization,
we transform the $e_g$ electron basis through the unitary transformation
$c_j=(a_{\alpha j}+i\,a_{\beta j})/\sqrt{2}$ and $d_j=(a_{\alpha j}-i\,a_{\beta j})/\sqrt{2}$\cite{koizumi}.
In the new basis the Hamiltonian (\ref{ham}) is rewritten as
\beq\label{hamtr}
H=&-&\sum_{\langle j,l \rangle}\left(\, t\,c_j^{\dagger}c_l^{\phantom{\dagger}}\,+\,t\,d_j^{\dagger}d_l^{\phantom{\dagger}}\,
+\,s\,c_j^{\dagger}d_l^{\phantom{\dagger}}\,+ s^*\,d_j^{\dagger}c_l^{\phantom{\dagger}}\right)  \\ 
&+& \sum_j\,
\left[\,V_j
\,c_j^{\dagger}d_j^{\phantom{\dagger}}\,+\,V_j^*\,d_j^{\dagger}c_j^{\phantom{\dagger}}
+\frac{U}{2}\,\langle n_j\rangle (c_j^{\dagger}c_j^{\phantom{\dagger}}+d_j^{\dagger}d_j^{\phantom{\dagger}}) \right. \nonumber \\
&+& \left. E_{JT}\left(q_{2j}^2+q_{3j}^2\right)-\frac{U}{4}\left(
\langle n_j\rangle^2 - \langle \tau_{z j}\rangle^2 - \langle \tau_{x j}\rangle^2\right)\right], \nonumber
\eeq
with $t=t_0/2$ and $s=\,e^{ i \phi_{\vert i-j\vert}}\,t_0/2$, the phase $\phi_{\vert i-j\vert}$ depending on the hopping
direction as $\phi_y=-\phi_x=\pi/3$. Therefore, the $e_g$ electrons pick up a phase change as they move
between different neighboring orbitals. The local
interaction $V_j=\vert
V_j\vert\,e^{i\xi_j}$ acquires a phase too \cite{koizumi}, being
\beq
\vert V_j\vert&=& 2E_{JT}\,\left[\bigl(\,q_{2j}- u\langle \tau_{xj}\rangle\bigr)^2 +
\bigl(q_{3j}-u\langle\tau_{zj}\rangle\bigr)^2\right]^{1/2}, \nonumber\\
\xi_j&=&\tan^{-1}\frac{q_{2j}-u \langle \tau_{xj}\rangle}{q_{3j}-u\langle\tau_{zj}\rangle},
\eeq
where we set $u=U/2E_{JT}$, while $\langle \tau_{\mu j} \rangle$ are the averaged pseudospin operators in the original basis
to be self-consistently determined.
As in undoped manganites there is only one $e_g$ electron per site, according to
Ref. [\onlinecite{koizumi}] we can take $q_{3j}=q_3$ and $q_{2j}=(-1)^j q_2$ [we numerically checked this assumption by
minimization of Hamiltonian (\ref{hamtr}) with respect to $q_{\alpha i}$]. The absolute value of the interaction
potential is then found to be constant,
while its phase changes within the chain with a periodicity equal to $\pi$. The Hamiltonian in momentum space can be
rewritten then in the electron basis $(c_k^\dagger,d_k^\dagger,c_{k+\pi}^\dagger,d_{k+\pi}^\dagger)$,
being $k$ defined, with the unit cell chosen as shown in Fig.\ref{fig1}(a), in the reduced Brillouin zone
$-\pi/2<k<\pi/2$:
\beq\label{hamk}
h_k=\left(\begin{array}{cc}
\mathcal{H}_{k,k} & \mathcal{H}_{k,k+\pi}\\
\mathcal{H}_{k,k+\pi} & \mathcal{H}_{k+\pi,k+\pi}\\
\end{array}
\right),
\eeq
where $\displaystyle{\mathcal{H}_{k,k+\pi}^{11}=\mathcal{H}_{k,k+\pi}^{22}=0}$, and
\beq
&\mathcal{H}_{k,k}^{11}&=\mathcal{H}_{k,k}^{22}=-t_0\cos k, \nonumber \\
&\mathcal{H}_{k,k}^{12}&=(\mathcal{H}_{k,k}^{21})^*=-\frac{t_0}{2}\bigl(\cos k_{+}+\cos k_{-}\bigr)\,+\,V_0,
\nonumber\\
&\mathcal{H}_{k,k+\pi}^{12}&=(\mathcal{H}_{k,k+\pi}^{21})^*=-\frac{t_0}{2}\bigl(\cos k_{-} - \cos k_{+} \bigr)\,+V_\pi, \nonumber \\
&\mathcal{H}_{k+\pi,k+\pi}^{11}&=\mathcal{H}_{k+\pi,k+\pi}^{22}=\,t_0\cos k, \nonumber \\
&\mathcal{H}_{k+\pi,k+\pi}^{12}&=(\mathcal{H}_{k+\pi,k+\pi}^{21})^*=\,\frac{t_0}{2}\bigl(\cos k_{+}+\cos
k_{-}\bigr)\,+\,V_0. \nonumber
\eeq

The hopping phase change enters in $k_\pm=k\pm\pi/3$, whereas the Fourier components of the local potential are 
$V_0=\vert V\vert \cos\xi_-\,e^{i\xi_+}$ and $V_\pi=\vert V\vert i\,\sin\xi_-\,e^{i\xi_+}$, being
$\xi_\pm=(\xi_1\pm\xi_2)/2$.
We note that the system is always in a band-insulating state, even when the JT coupling and $U$ are set to zero. The band
insulator is stabilized by the
phase difference between the interorbital hopping amplitudes, as it happens in the AFM-CE phase of half-doped
manganites\cite{dagotto.rev,brink}, and it is very robust due to its topological origin.
However, the OO emerges only when the JT and Coulomb interactions are sufficiently strong to induce a
nonzero $\vert V\vert$.
In order to inspect the OO in real space, we can evaluate the local average orbital occupancy $\rho_{\gamma j}=\langle \phi_{\gamma j}^\dagger
\phi_{\gamma j}^{\phantom{\dagger}}\rangle$, where $\phi_{\gamma j}=-\sin(\theta_{\gamma j}/2)\,a_{\alpha j}+
\cos(\theta_{\gamma j}/2)\,a_{\beta j}$
and $\theta_{\gamma i}=2\pi/3 \,(4\pi/3)$ for $\gamma=3x^2-r^2\,(3y^2-r^2)$. In Fig.\ref{fig1}(c) we show the evolution
of $\rho_{3x^2-r^2}$ on site 1 (equivalent to $\rho_{3y^2-r^2}$ on site 2) as a function of JT interaction at $U=0$
and the OO pattern of alternating
$3x^2-r^2/3y^2-r^2$ within the chain. We note that a similar result is obtained for
the linear chain, where the hopping direction and the related phase $\phi_{\vert i-j\vert}$ are unchanged,
while the OO-induced phase change is retained\cite{koizumi}:
in this case the band insulator has no topological origin and
emerges in an orbitally ordered pattern because of the local interaction, being metallic for
$\vert V\vert=0$ [dotted line in Fig.\ref{fig1}(c)].  On the other hand, the inclusion of $U$, albeit at a mean-field
level, makes the OO onset more robust [Fig\ref{fig1}(d)].

We consider now the possible ferroelectricity. Following the prescription described in Ref. [\onlinecite{resta2}], we evaluate the polarization as
\beq\label{po}
P=-\frac{e}{2\pi}\lim_{L\to\infty}\mbox{Im}\,\ln\,\det S,
\eeq
where $L$ is the chain length (periodic boundary conditions are assumed) and $S$ is the overlap matrix, defined as
$
S_{m,m^{\prime}}(k,k^{\prime})=\langle \psi_{m,k}\vert e^{-i\frac{2\pi}{L}\hat{x}}\vert \psi_{m^{\prime},k^{\prime}}\rangle.
$
The eigenvectors $\psi_{m,k}$, where $m$ is a band index, can be expressed as
\beq
\vert \psi_{k,m}\rangle =&& \left(U_k(m,1)\,c^\dagger_{k} + U_k(m,2)\,d^\dagger_{k} + \right. \nonumber\\
&&\left. U_k(m,3)\,c^\dagger_{k+\pi} + U_k(m,4)\,d^\dagger_{k+\pi}\right)\,\vert 0\rangle.
\eeq
Here $U_k$ is the unitary matrix which diagonalizes the Hamiltonian (\ref{hamk}) at each $k$, and $\vert 0\rangle$ is the vacuum state.
By defining the position operator as $\hat{x}=\sum_j\,j\,(c^\dagger_j\,c^{\phantom{\dagger}}_j+
d^\dagger_j\,d^{\phantom{\dagger}}_j)$, one sees that $S$ elements vanish except when each pair of vectors
$k,k^{'}$ differs by an amount $\varepsilon=2\pi/L$; then the determinant can be factorized into $L$ small determinants
whose dimension is equal to the number of occupied bands \cite{resta2}, giving $\det S=\Pi_{k} \det S(k,k+\varepsilon)$,
with the small overlap matrix being
\beq\label{so}
S_{m,m^{\prime}}(k,k+\varepsilon)=\sum_{\gamma=1,4}U_{k}^\dagger(\gamma,m)U_{k+\varepsilon}^{\phantom{\dagger}}(m^\prime,\gamma).
\eeq

\begin{figure}
\includegraphics[width=7cm]{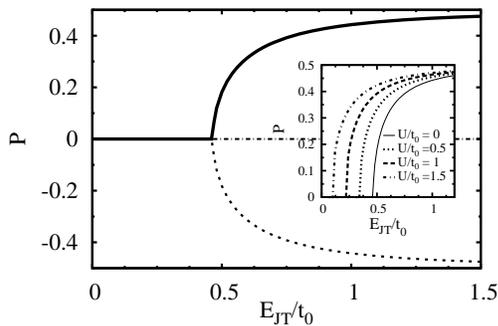}
\caption{Polarization in the one-dimensional zigzag chain as a function of $E_{JT}$; the electron charge $e$ is taken equal
to unity. Solid (dashed) line is evaluated for the $3x^2-r^2/3y^2-r^2$ ($3y^2-r^2/3x^2-r^2$) OO pattern; dot-dashed line is
evaluated within the linear chain. Inset: polarization for different values of $U$.}\label{pol}
\end{figure}

In Fig.\ref{pol} we show the polarization evaluated through formula (\ref{po})
for the ground state of our model: 
it remains equal to zero in the insulating phase as far
as no OO is induced in the zigzag chain by the local interaction, then it rapidly increases,
closely following the evolution of the occupied orbital density. 
The sign of
$P$ can be changed by rotating the OO pattern, in such a way that orbital state
$3y^2-r^2\,(3x^2-r^2)$ is
occupied at site 1(2) instead of $3x^2-r^2\,(3y^2-r^2)$.
It is worthwhile to notice that $P$ is always zero 
in the orbital-ordered insulating phase found in the
linear chain, even though the interaction phase $\xi_j$ has been shown to induce a Berry phase in the electronic Bloch
function for a given $k$\cite{koizumi}. Indeed, this geometric phase takes into account the difference in the electron
motion, clockwise or counterclockwise, around each site displaying OO within the linear chain; however, this difference
sums up to zero when evaluated along the
whole linear chain. Similarly, when the local interaction is not strong enough to induce OO along the zigzag chains, the
difference in the electron motion due to
direction-dependent interorbital hopping amplitudes, which induces a phase change in the electronic Bloch functions, does
not give rise alone to any nonvanishing $P$. On the other hand, our result suggests that {\em the
interplay between the phase changes induced by the OO and direction-dependent hopping amplitudes} is
responsible for the onset of a ferroelectric state, which has ultimately a topological origin. 
We stress that this polarization has a purely electronic origin, since
the position of the ions is fixed within the chains,
supporting the DFT calculations which report a large electronic contribution to $P$ in the whole class of
AFM-E $o-R$MnO$_3$\cite{yamauchi}. We can estimate the magnitude of the calculated electronic $P$ for a
realistic
three-dimensional system multiplying it by
a scale factor $a_0/V_0$, where $a_0$ is the Mn-Mn distance (corresponding to the lattice constant in our simplified
model) and $V_0$ is the unit cell volume. We estimate $a_0=3.89-4\,$ \AA~ and $V_0=244-220$ \AA$^3$
for $R$MnO$_3$\cite{alonso},
implying $P_{el}$ of the order of  $\sim 10\, \mu C/cm^2$. On the other hand, both $E_{JT}/t_0$ and $U/t_0$
are of the order of unity in manganites, being $t_0=0.1-0.5\; eV$, $E_{JT}\simeq 0.25\; eV$ and
$U\gtrsim1\; eV$\cite{dagotto.rev}, suggesting
that the OO found in our simple model is easily realized in undoped manganites.

To get more physical insight, we evaluated the position of the WF centers in the chain.
For this purpose, the construction of
maximally localized WFs is not needed in one-dimensional systems.
Indeed, they can be obtained as the eigenvalues of a matrix $\Lambda$, constructed as the product of the unitary
parts of the $S$ matrices along the $k$-point string (by ``unitary part''  we mean the matrix product $VW^\dagger$ taken
from the singular value decomposition $S=V\Sigma W^\dagger$, where $V$ and $W$
are unitary and $\Sigma$ is a diagonal matrix with nonnegative
diagonal elements)\cite{marzari}. We find that i) in the insulating phase with no OO, the WFs are bond centered, their
centers located
exactly in the middle of each bond connecting two neighboring sites along the zigzag
chain, and
ii) when OO is established, WF centers move (in the same direction along the zigzag
chain) toward lattice sites. This unveils the reduced symmetry of the system, that loses the center of
symmetry in the middle of each bond, thus allowing for a finite $P$. The displacement
direction is related to the character of the occupied orbitals,
in the sense that each WF center moves from the middle of the bond toward the
neighboring site where the occupied orbital is aligned parallel to the vector connecting
each pair of sites. Since the OO pattern is $3x^2-r^2/3y^2-r^2$ (or
$3y^2-r^2/3x^2-r^2$)  and the direction along the chain changes as $\{..x,y,x,y..\}$, there is a net "leftward" (or
"rightward") displacement of WF centers.

Because of the analogies shared with the AFM-CE phase relevant for half-doped manganites, showing double zigzag spin
chains\cite{dagotto.rev}, let us briefly discuss the outcome of
our analysis in that case. The Hamiltonian (\ref{hamtr}) is still suitable to describe the motion of $e_g$ electrons
within chains where the hopping direction changes as $\{..x,x,y,y..\}$, as expected on the background of the ferromagnetically aligned
core-spin chains characteristic of the AFM-CE configuration. We can then distinguish between corner Mn sites, where hopping
changes direction, and bridge Mn sites, where it does not\cite{brink}. As pointed out before, the system is a band insulator
even in the absence of any interaction, due to the effective dimerization induced by the phase change in the
hopping amplitudes at corner sites. However, 
an OO pattern of $3x^2-r^2/3y^2-r^2$ states already
occurs on bridge sites\cite{dagotto.rev,brink}, at odds with the AFM-E phase considered so far, where only corner
sites appear. Turning on the local interaction gives rise to a charge transfer
from corner to bridge sites, but leaves the OO pattern unchanged.
We can evaluate  $P$ along the same lines described before.  We notice however that electrons pick up a phase on
{\em corner} sites, depending on the change of hopping direction, and a different one on {\em bridge} sites,
related to the orbital-ordered state, {\em which do not interfere with each other}.
As a result, we find that $P$ is always zero with a CE-type constraint on the electron motion.

We thank Professor D. Vanderbilt for reading the manuscript and for his helpful comments.
P.B. would like to thank Professor Y. Mokrousov and Dr. G. D'Avino for useful and fruitful discussions.
This work has been supported by the European Community's
Seventh Framework Programme FP7/2007-2013 under
Grant agreement No. 203523-BISMUTH.

\end{document}